\documentstyle[12pt,epsf]{article}

\begin{document}

\baselineskip 6mm
\renewcommand{\thefootnote}{\fnsymbol{footnote}}


\newcommand{\nc}{\newcommand}
\newcommand{\rnc}{\renewcommand}


\rnc{\baselinestretch}{1.24}    
\setlength{\jot}{6pt}       
\rnc{\arraystretch}{1.24}   

\makeatletter
\rnc{\theequation}{\thesection.\arabic{equation}}
\@addtoreset{equation}{section}
\makeatother



\nc{\be}{\begin{equation}}

\nc{\ee}{\end{equation}}

\nc{\bea}{\begin{eqnarray}}

\nc{\eea}{\end{eqnarray}}

\nc{\xx}{\nonumber\\}

\nc{\ct}{\cite}

\nc{\la}{\label}

\nc{\eq}[1]{(\ref{#1})}

\nc{\newcaption}[1]{\centerline{\parbox{6in}{\caption{#1}}}}

\nc{\fig}[3]{

\begin{figure}
\centerline{\epsfxsize=#1\epsfbox{#2.eps}}
\newcaption{#3. \label{#2}}
\end{figure}
}


\def\CA{{\cal A}}
\def\CC{{\cal C}}
\def\CD{{\cal D}}
\def\CE{{\cal E}}
\def\CF{{\cal F}}
\def\CG{{\cal G}}
\def\CH{{\cal H}}
\def\CK{{\cal K}}
\def\CL{{\cal L}}
\def\CM{{\cal M}}
\def\CN{{\cal N}}
\def\CO{{\cal O}}
\def\CP{{\cal P}}
\def\CS{{\cal S}}
\def\CU{{\cal U}}
\def\CW{{\cal W}}
\def\CY{{\cal Y}}


\def\IR{{\hbox{{\rm I}\kern-.2em\hbox{\rm R}}}}
\def\IB{{\hbox{{\rm I}\kern-.2em\hbox{\rm B}}}}
\def\IN{{\hbox{{\rm I}\kern-.2em\hbox{\rm N}}}}
\def\IC{\,\,{\hbox{{\rm I}\kern-.59em\hbox{\bf C}}}}
\def\IZ{{\hbox{{\rm Z}\kern-.4em\hbox{\rm Z}}}}
\def\IP{{\hbox{{\rm I}\kern-.2em\hbox{\rm P}}}}
\def\IH{{\hbox{{\rm I}\kern-.4em\hbox{\rm H}}}}
\def\ID{{\hbox{{\rm I}\kern-.2em\hbox{\rm D}}}}


\def\a{\alpha}
\def\b{\beta}
\def\ga{\gamma}
\def\d{\delta}
\def\ep{\epsilon}
\def\ph{\phi}
\def\k{\kappa}
\def\l{\lambda}
\def\m{\mu}
\def\n{\nu}
\def\th{\theta}
\def\rh{\rho}
\def\s{\sigma}
\def\t{\tau}
\def\w{\omega}
\def\G{\Gamma}


\def\half{\frac{1}{2}}
\def\dint#1#2{\int\limits_{#1}^{#2}}
\def\goto{\rightarrow}
\def\para{\parallel}
\def\brac#1{\langle #1 \rangle}
\def\grad{\nabla}
\def\curl{\nabla\times}
\def\div{\nabla\cdot}
\def\p{\partial}
\def\e{\epsilon_0}


\def\Tr{{\rm Tr}\,}
\def\det{{\rm det}}


\def\vare{\varepsilon}
\def\bz{\bar{z}}
\def\bw{\bar{w}}


\def\do{{\bf R}_{NC}^{4}}
\def\re{{\bf R}_{NC}^2}
\def\mi{{\bf R}_C^2}
\def\c{{\bf C}}
\def\z{{\bf Z}}

\begin{titlepage}

\hfill\parbox{5cm} {SOGANG-HEP 294/02}

\hfill\parbox{4cm} {{\tt hep-th/0205010}}

\vspace{25mm}

\begin{center}
{\Large \bf Zero Modes and the Atiyah-Singer Index 
\\ in Noncommutative Instantons}

\vspace{15mm}
Keun-Young Kim$^a$\footnote{kykim@physics4.sogang.ac.kr}, Bum-Hoon
Lee$^a$\footnote{bhl@ccs.sogang.ac.kr} and Hyun Seok
Yang$^b$\footnote{hsyang@phys.ntu.edu.tw}
\\[10mm]

$^a${\sl Department of Physics, Sogang University,
Seoul 121-742, Korea} \\
$^b${\sl Department of Physics, National Taiwan University,
Taipei 106, Taiwan, R.O.C.} \\
\end{center}

\thispagestyle{empty}

\vskip2cm


\centerline{\bf ABSTRACT} 
\vskip 4mm 
\noindent

We study the bosonic and fermionic zero modes 
in noncommutative instanton backgrounds 
based on the ADHM construction. In $k$ instanton background 
in $U(N)$ gauge theory, we show how to explicitly construct $4Nk \; (2Nk)$ 
bosonic (fermionic) zero modes in the adjoint representation 
and $2k \; (k)$ bosonic (fermionic) zero modes in the fundamental
representation from the ADHM construction. 
The number of fermionic zero modes 
is also shown to be exactly equal to the Atiyah-Singer index 
of the Dirac operator in the noncommutative instanton background. 
We point out that (super)conformal zero modes in non-BPS 
instantons are affected by the noncommutativity. 
The role of Lorentz symmetry breaking by the noncommutativity is also
briefly discussed to figure out the structure of $U(1)$ instantons. \\

PACS numbers: 11.15.-q, 11.15.Tk, 02.40.Gh 

\vspace{1cm}

\today

\end{titlepage}

\renewcommand{\thefootnote}{\arabic{footnote}}
\setcounter{footnote}{0}

\section{Introduction}

Instantons were found by Belavin, Polyakov, Schwartz and Tyupkin
(BPST) \ct{bpst} almost thirty years ago, as topologically nontrivial
solutions of the duality equations of the Euclidean Yang-Mills theory 
with finite action. Immediately instantons were realized to describe
the tunnelling processes between different $\theta$-vacua in Minkowski
space and lead to the strong CP problem in QCD 
\ct{jare,cdg}. 
(For the earlier development of instanton physics, see the collection
of papers \ct{shifman}.) The non-perturbative chiral anomaly in the
instanton background led to baryon number violation and a solution to
the $U(1)$ problem \ct{thooft1,thooft2}. These revealed that instantons can
have their relevance to phenomenological models like QCD and the
Standard model \ct{ss}.

Instanton solutions also appear as BPS states in string theory. 
They are described by Dp-branes bound to D(p+4)-branes 
\ct{witten,douglas}. Subsequently, in \ct{sv,cm}, low-energy
excitations of D-brane bound states were used to explain the 
microscopic degrees of freedom of black-hole entropy, for which the
information on the instanton moduli space has a crucial role. In
addition the multi-instanton calculus was used for a nonperturbative test
of AdS/CFT correspondence \ct{bg,bgkr,dhkmv,bnv}, where 
the relation between Yang--Mills instantons and D-instantons was 
beautifully confirmed by the explicit form of the classical
D-instanton solution in $AdS_5 \times {\bf S}^5$ background 
and its associated supermultiplet of zero modes.

Recently instanton solutions on noncommutative spaces have been turned out 
to have more richer spectrums. While commutative instantons are always
BPS states, noncommutative instantons admit both BPS and non-BPS states.
Especially, instanton solutions can be found in $U(1)$ gauge theory 
and the moduli space of non-BPS instantons is smooth, 
small instanton singularities being resolved 
by the noncommutativity \ct{ns,sw}. Remarkably, instanton solutions in
noncommutative gauge theory can also be studied by
Atiyah-Drinfeld-Hitchin-Manin (ADHM) equation \ct{adhm} slightly
modified by the noncommutativity \ct{ns}. 
ADHM construction uses some quadratic matrix equations, hence 
noncommutative objects in nature, to construct
(anti-)self-dual configurations of the gauge field.
Thus the noncommutativity of space doesn't make any serious obstacle
for the ADHM construction of noncommutative instantons and indeed it
turns out that it is a really powerful tool even for noncommutative
instantons. Recently much progress has been made in this direction 
\ct{ns,sw,ly,bn,kf1,kko,ho,kly1,kf2,lty,kf3,
asch,corr,ckt,ham,kuro,kly2,lp,iks}.

This paper is aiming to explain how to construct bosonic and
fermionic zero modes in noncommutative instanton backgrounds 
based on the ADHM construction and how to relate them 
to the Atiyah-Singer index. Recently several papers 
\ct{lty,dhk1,hkt,dhk2,klyi,holl} discussed the
instanton moduli space and the instanton calculus in noncommutative spaces.  
This paper is organized as follows. In next section we briefly review
the ADHM construction on noncommutative spaces and explain how to
construct the zero modes in noncommutative instanton backgrounds 
from the ADHM construction. In Section 3, we discuss the moduli space
$\CM_{k,N}$ of $k$ instantons in noncommutative $U(N)$ gauge theory
and how to explicitly construct fermion zero modes in adjoint and fundamental
representations. In $U(1)$ instanton background 
the fermionic zero modes in the fundamental
representation were briefly discussed by Nekrasov \ct{nekra} recently,
which are precisely reproduced from our solution for the $U(1)$ case. 
We point out that (super)conformal
zero modes in the non-BPS background are affected 
by the noncommutativity and 
their explicit construction may be nontrivial, calling for further study.
We also speculate that the $U(1)$ instanton may be understood by the
structure of Lorentz symmetry breaking by the noncommutativity.
In Section 4, we show that the number of the bosonic and the fermionic 
zero modes constructed in Section 3 is related to the Atiyah-Singer
index of the noncommutative instantons. In Section 5 we discuss the
results obtained and address some issues.

\section{Zero Modes in Instantons}

In this section we review briefly the formalism of the ADHM
construction for instantons and discuss how to find the zero modes around
the instanton background from this formalism 
\ct{bcgw,cfgt,osborn,csw,cgt,cg}. Here the ADHM construction
is universally valid both for commutative and for noncommutative spaces.
We will specify the noncommutative case if necessary. Although we
discuss both self-dual and anti-self-dual instantons,
we will often consider the anti-self-dual instantons as a specific
problem. The self-dual instantons could be treated similarly.

Our interest is how to construct all the finite action solutions of
the (anti-)self-duality equation in $U(N)$ gauge theory
\begin{equation}\label{self-dual}
F_{\mu\nu}=\pm *F_{\mu\nu}=\pm \half \vare_{\mu\nu\rho\sigma}F_{\rho\sigma}.
\end{equation}
where the field strength $F_{\mu\nu}$ is given by
\begin{equation} \la{fsF}
F_{\mu\nu}=\partial_\mu A_\nu-\partial_\nu A_\mu
+[A_\mu,A_\nu].
\end{equation}
In noncommutative space, for the field strength $F_{\mu\nu}$ 
to be gauge covariant, 
we need the commutator term in \eq{fsF} even for $U(1)$ gauge group.
ADHM construction provides an algebraic way for constructing
(anti-)self-dual configurations of the gauge field in terms of some
quadratic matrix equations on four manifolds \ct{adhm}.

In order to discuss the ADHM construction, it is convenient to
introduce quaternions defined by
\begin{equation}\label{quaternion}
  {\bf x}=x_\mu \sigma^\mu, \qquad {\bar{\bf x}}=x_\mu {\bar\sigma}^\mu,
\end{equation}
where $\sigma^\mu=(i \tau^a, 1)$ and ${\bar \sigma}^\mu=(-i
\tau^a, 1)$ which have the basic properties 
\bea \la{sigma} 
&& \sigma^\mu {\bar \sigma}^\nu=\delta^{\mu\nu}+i\sigma^{\mu\nu},
\qquad \sigma^{\mu\nu}=\eta^a_{\mu\nu} \tau^a=*\sigma^{\mu\nu},\xx
&& {\bar \sigma}^\mu \sigma^\nu=\delta^{\mu\nu}+i{\bar
\sigma}^{\mu\nu}, \qquad {\bar \sigma}^{\mu\nu}={\bar
\eta}^a_{\mu\nu} \tau^a= -*{\bar \sigma}^{\mu\nu}. 
\eea 
The $\sigma^\mu$ and ${\bar \sigma}^\mu$ can be used to construct the
Euclidean Dirac matrices as 
\bea \la{gamma} 
&& \gamma^\mu =
\left(\begin{array}{cc}
0 & {\bar \sigma}^\mu \\
\sigma^\mu & 0 \end{array}\right), \qquad
\gamma_5 = \gamma_1 \gamma_2 \gamma_3 \gamma_4 =\left(\begin{array}{cc}
1 & 0 \\
0 & -1 \end{array}\right),\xx
&& \{ \gamma^\mu, \gamma^\nu \}=2 \delta^{\mu\nu}, \qquad
[\gamma^\mu, \gamma^\nu] = 2i\left(\begin{array}{cc}
{\bar \sigma}^{\mu\nu} & 0 \\
0 & \sigma^{\mu\nu} \end{array}\right).
\eea

In the ADHM construction the gauge field with instanton number $k$
for $U(N)$ gauge group is given in the form
\begin{equation}\label{A}
 A_\mu (x)=v(x)^{\dagger} \partial_\mu v(x)
\end{equation}
where $v(x)$ is $(N+2k)\times N$ matrix defined by
the equations \footnote{One can always normalize the matrix $v$ in usual 
way even for noncommutative $\do$ and $\re \times \mi$ 
as done in \ct{ckt,kly2}, respectively.}
\begin{eqnarray}
\label{normalization}
  && v(x)^\dagger v(x)=1,\\
\label{zero-mode} 
&& v(x)^\dagger \Delta(x)=0.
\end{eqnarray}
In (\ref{zero-mode}), $\Delta(x)$ is a $(N+2k) \times 2k$ matrix, 
linear in the position variable $x$, having the
structure
\begin{equation}\label{Delta}
  \Delta (x)= \left \{ \begin{array}{l} a - b {\bf x},\qquad \mbox{
  self-dual instantons},\\
 a - b {\bar {\bf x}},\qquad \mbox{
  anti-self-dual instantons},
\end{array} \right.
\end{equation}
where $a, b$ are $(N+2k) \times 2k$ matrices. $\Delta(x)$ can be
thought of as a map from a $2k$-complex dimensional space $W$ to a
$N+2k$-complex dimensional space $V$. The matrices $a, b$ are
constrained to satisfy the conditions that $\Delta(x)^\dagger
\Delta(x)$ be invertible and that it commutes with the
quaternions. These conditions imply that $\Delta(x)^\dagger
\Delta(x)$ as a $2k \times 2k$ matrix has to be factorized as
follows 
\be \la{invertible}
\Delta(x)^\dagger \Delta(x) =f^{-1}(x) \otimes  1_2
\ee 
where $f^{-1}(x)$ is a $k \times k$ matrix and $1_2$ is a unit
matrix in quaternion space. Then the resulting field strength
$F_{\mu\nu}$ obtained by \eq{A} ensures the (anti-)self-duality equation
\eq{self-dual} if and only if $v(x)$ and $\Delta(x)$ obey the
completeness relation
\begin{equation} \la{complete}
v(x) v(x)^\dagger + \Delta(x) f(x) \Delta(x)^\dagger =1.
\end{equation}
Note that the matrix $v(x)$ in (\ref{normalization}) and (\ref{zero-mode}) 
is unique only up to a gauge transformation
\be \la{un}
v(x) \rightarrow v(x) g(x), \qquad  
\Delta (x) \rightarrow \Delta (x), \qquad g(x) \in U(n)
\ee
which generates usual $U(N)$ gauge transformations for the gauge field
\be \la{gaugetr}
A_\mu (x) \rightarrow g(x)^\dagger A_\mu (x) g(x)+ g(x)^\dagger \p_\mu
g(x).
\ee

Given a pair of matrices $a, b$, (\ref{normalization}) and
(\ref{zero-mode}) define $A_\mu$ up to gauge equivalence.
Different pair of matrices $a, b$ may yield gauge equivalent
$A_\mu$ since (\ref{normalization}) and (\ref{zero-mode}) are
invariant under
\begin{equation}\label{ugl}
  a \rightarrow QaK, \quad b \rightarrow QbK,  \quad v \rightarrow Qv
\end{equation}
where $Q \in U(N+2k)$ and $K \in GL(k,\c)$. This freedom can be
used to put $a, b$ in the canonical forms 
\be \la{ab} 
a = \left(\begin{array}{c}
\lambda \\
\xi \end{array}\right), \qquad b=\left(\begin{array}{c}
0 \\
1_{2k} \end{array}\right), 
\ee
where $\lambda$ is an $N \times 2k$ matrix and $\xi$ is a $2k \times
2k$ matrix. Here we decompose the matrix $\xi$ in the quaternionic
basis ${\bar \sigma}^\mu$ as a matter of convenience
\begin{equation}
\xi = \xi_\mu {\bar \sigma}^\mu, 
\end{equation}
where $\xi_\mu$'s are $k \times k$ matrices. In the basis \eq{ab}, the
constraint \eq{invertible} boils down to
\bea \la{adhm1}
&& {\rm tr}_2 \tau^a a^\dagger a = 
\left \{ \begin{array}{l} \theta^{\mu\nu}{\bar \eta}^a_{\mu\nu},
\qquad \mbox{self-dual instantons},\\
 \theta^{\mu\nu}\eta^a_{\mu\nu},\qquad \mbox{anti-self-dual instantons},
\end{array} \right. \\
\la{adhm2}
&& \xi_\mu^\dagger = \xi_\mu,
\eea
where ${\rm tr}_2$ is the trace over the quaternionic indices.
Here we work in general in flat noncommutative Euclidean space 
${\bf R}^4$ represented by
\begin{equation}\label{NC-space}
  [x^\mu,x^\nu]=i\theta^{\mu\nu}
\end{equation}
where $\theta^{\mu\nu}=-\theta^{\nu\mu}$.
Note that there still exists a residual $U(k)$ symmetry of 
the $U(N+2k) \times GL(k, \c)$ symmetry group \eq{ugl} to preserve 
the canonical form for $b$ given in \eq{ab} 
which acts on $\lambda$ and $\xi_\mu$ as
\begin{equation} \la{uk}
\lambda \rightarrow \lambda U, \qquad 
\xi_\mu \rightarrow U^\dagger \xi_\mu U, \qquad U \in U(k).
\end{equation}

Introduce two $N \times k$ matrices $I^\dagger,\,J$ 
to represent the $N \times 2k$ matrix $\lambda$ as 
\be \la{lambda}
\lambda =(I^\dagger, J).
\ee
Then the ADHM constraints \eq{adhm1} for the
anti-self-dual instantons, for example, have the following forms
\bea \la{adhm3}
&& \mu_r = [B_1,B_1^\dagger]+[B_2,B_2^\dagger]+II^\dagger-J^\dagger J
=\theta^{\mu\nu}\eta^3_{\mu\nu} \\
\la{adhm4}
&& \mu_c = IJ+[B_1,B_2]=\half \theta^{\mu\nu}
(\eta^1_{\mu\nu}+i \eta^2_{\mu\nu}),
\eea
where $B_1=\xi_2+i\xi_1,\,B_2=\xi_4+i\xi_3$. 
Now we can count the independent bosonic collective coordinates of
multi-instanton solution. From the ADHM construction for $k$
instantons in $U(N)$ gauge theory, the matrices $\lambda$ and $\xi_\mu$
in \eq{ab} have $4Nk + 4k^2$ real degrees of freedom. However
\eq{adhm1} imposes $3k^2$ constraints for $\lambda$ and $\xi_{\mu}$. 
In addition we have the $U(k)$ residual symmetry \eq{uk}. 
These remove $4k^2$ degrees of freedom and thus totally $4Nk$
real degrees of freedom remain. \footnote{This contains the global
gauge rotations satisfying \eq{gauge} which are non-normalizable 
on ${\bf R}^4$. To obtain purely physical zero modes, the global gauge
rotations are removed, leaving the number of physical zero modes as
$4Nk-N^2+1$ for $k \ge {N \over 2}$, and $4k^2 +1$ for $k \le {N \over
  2}$ \ct{bcgw,cfgt,osborn,csw}. 
However the global gauge rotations appear in the $k$-instanton
measure in the functional integral and in the calculation of the
Atiyah-Singer index. So we will keep these modes.} 
They describe a $4Nk$-dimensional
moduli space $\CM_{k,N}$ which is still hyper-K\"ahler space. 
The ADHM constraints \eq{adhm3} and \eq{adhm4} 
are nothing but the (deformed) moment maps of the hyper-K\"ahler quotient
construction. We will show that even for noncommutative spaces 
$\CM_{k,N}$ is the moduli space of $k$ instantons in $U(N)$ gauge theory.

Let's consider Yang-Mills theory with gauge group $U(N)$ with action
\be \la{ym-action}
S=-\half \int d^4 x {\rm tr} F_{\mu\nu} F_{\mu\nu}.
\ee
Of course, for the noncommutative space \eq{NC-space}, the integral in
\eq{ym-action} may be replaced by an appropriate trace over 
the representation space $\CH$ of the algebra \eq{NC-space}:
\be
\int d^4 x \quad \rightarrow \quad \Tr_\CH.
\ee 
Consider small fluctuations about a classical instanton
solution satisfying \eq{self-dual}
\be \la{gauge-fluc}
A_\mu(x) = A_\mu^{cl}(x) + \delta A_\mu(x).
\ee
Since we are only interested in {\it physical} excitations, 
the fluctuations $\delta A_\mu (x)$ are required to be
square integrable \footnote{For noncommutative spaces, this condition 
means that $\Tr_\CH |\delta A_\mu (x)|^2  < \infty$.}
and to be orthogonal to gauge transformations, i.e.
\be \la{ortho}
\int d^4x  {\rm tr} D_\mu \Lambda(x) \delta A_\mu(x)=
-\int d^4x  {\rm tr} \Lambda(x) D_\mu \delta A_\mu(x)=0
\ee
where $\Lambda(x)$ is an arbitrary gauge parameter in $U(N)$. 
Since the condition \eq{ortho} 
should be satisfied for any $\Lambda(x)$, this orthogonality requirement
is equivalent to the usual gauge condition
\be \la{gauge}
D_\mu \delta A_\mu (x)=0.
\ee

If the action is expanded to second order in $\delta A_\mu$, 
one can find the following result 
\be \la{vary-action}
S[A_\mu] \approx S[A_\mu^{cl}] -\frac{1}{4} \int d^4 x 
{\rm tr} (\delta F_{\mu\nu}
\mp * \delta F_{\mu\nu})^2
\ee
where $\delta F_{\mu\nu}=D_\mu \delta A_\nu - D_\nu \delta A_\mu$ with 
$D_\mu \delta A_\nu =
\p_\mu \delta A_\nu +[A_\mu^{cl}, \delta A_\nu]$. To derive the result
\eq{vary-action} \footnote{In the noncommutative case, one may use 
the cyclic property under the integral $\Tr_\CH$.} 
we have used the (anti-)self-duality condition \eq{self-dual} and 
the equations of motion for $A_\mu^{cl}$ 
and the fact that the boundary contribution at infinity
is trivial since the fluctuations $\delta A_\mu (x)$ are 
square integrable. The above result shows that as long as the
fluctuations, satisfying the gauge condition \eq{gauge}, 
around a classical instanton solution $A_\mu^{cl}(x)$
satisfy the (anti-)self-duality equation,
\be \la{zero-self}
\delta F_{\mu\nu}=\pm * \delta F_{\mu\nu},
\ee
they are zero modes, i.e. the action is not changed by these
collective excitations. Thus one can study the space of
(anti-)self-dual solutions by considering small fluctuations 
$A_\mu^{cl}(x) + \delta A_\mu(x)$ about a particular solution 
$A_\mu^{cl}(x)$ and asking that the resulting field strength continue
to be (anti-)self-dual.

Now our problem is how to find the fluctuations, the zero modes, 
satisfying \eq{zero-self} and \eq{gauge}. One strategy to characterize 
the general family of instanton solutions by all relevant collective 
coordinates is following \ct{jnr}. One starts from some known solution and
transforms it by applying all symmetry transformations. The
symmetry transformations which act nontrivially generate new
solutions and require an introduction of the collective
coordinates. For example, for commutative $SU(2)$ instantons, 
these are positions $x_0^\mu$ by translation
symmetry, size $\rho$ by dilatation and three orientation
parameters $\omega^a$ corresponding to global $SU(2)$ rotations.
However this method is not easy to generalize to noncommutative spaces
although it may be interesting.

Instead we will use the ADHM construction 
to find the zero modes. As discussed in this section,
the matrices $a, b$ carry full information for collective coordinates 
up to the gauge transformation (\ref{ugl}).  
Consider so varying $\Delta(x)$ by an amount $\delta
\Delta(x)=\delta a - \delta b {\bf x}$ 
(or $\delta a - \delta b {\bar {\bf x}}$) 
within the space of solutions.
$\delta \Delta$ is given
by $\delta a$ and $\delta b$ which describe collective variables
in the moduli space $\CM_{k,N}$.  After the gauge fixing
\eq{ab}, all the collective coordinates are encoded in the matrix $a$
up to the residual $U(k)$ symmetry \eq{uk}. Here we will use the gauge
fixed basis \eq{ab} and thus we will put $\delta b=0$. 
Then $v(x)$ will have to be
changed to $(v+\delta v)(x)$ where
\begin{equation}\label{variation1}
  \delta v^\dagger \Delta + v^\dagger \delta \Delta=0
\end{equation}
and
\begin{equation}\label{variation2}
  \delta v^\dagger v + v^\dagger \delta v =0.
\end{equation}
The general solution to (\ref{variation1}) is of the form \ct{cgot}
\begin{equation}\label{sol-var1}
  \delta v^\dagger = - v^\dagger \delta \Delta f \Delta^\dagger + \delta
  u^\dagger v^\dagger
\end{equation}
where $\delta u= v^\dagger \delta v$ and (\ref{variation2})
implies
\begin{equation}\label{sol-var2}
  \delta u + \delta u^\dagger=0.
\end{equation}
Since we are still in the space of solutions, $(\Delta+\delta
\Delta)(x)$ has to satisfy the factorization condition \eq{invertible}
and the completeness relation \eq{complete}. By using \eq{sol-var1} and
\eq{sol-var2}, one can easily check that $(\Delta+\delta
\Delta)(x)$ automatically satisfies the completeness relation. Since
$\delta \Delta=\delta a$, this variation is equivalent to 
$a \rightarrow a + \delta a$. Thus, from \eq{adhm1} and \eq{adhm2}, 
we see that $\delta a$ should satisfy
\be \la{zero-adhm}
\delta \mu^a = {\rm tr}_2 \tau^a (\delta a^\dagger a 
+ a^\dagger \delta a)=0
\ee
or explicitly 
\bea \la{zero-adhm1}
&& \delta \mu_r = [\delta B_1,B_1^\dagger]+[B_1, \delta B_1^\dagger]+ 
 [\delta B_2, B_2^\dagger]+[B_2, \delta B_2^\dagger]\xx
&& \qquad \;\; +\delta II^\dagger + I \delta I^\dagger- \delta J^\dagger J
-J^\dagger \delta J =0, \\
\la{zero-adhm2}
&& \delta \mu_c = \delta IJ+ I\delta J
+[\delta B_1, B_2]+[B_1,\delta B_2]=0.
\eea
One sees that the ADHM constraints for the variation $\delta a$
are never deformed by the noncommutativity and are completely
equivalent to their commutative counterparts. However, we will see
later that some zero modes are still affected by the noncommutativity
since they are represented by the original moduli $a$ which should
satisfy the constraints \eq{adhm1}.

From the solution \eq{sol-var1}, 
\begin{eqnarray}\label{varA}
\delta A_\mu &=& v^\dagger \partial_\mu \delta v + \delta
v^\dagger \partial_\mu v \nonumber \\
&=& v^\dagger (\delta \Delta f \partial_\mu \Delta^\dagger -
\partial_\mu \Delta f \delta \Delta^\dagger) v + D_\mu \delta u.
\end{eqnarray}
The term $D_\mu \delta u = \partial_\mu \delta u + [A_\mu^{cl}, \delta
u]$ corresponds to an infinitesimal gauge transformation. 
By a suitable choice of gauge for $A_\mu^{cl}$, we will put $\delta
u=0$. \footnote{Using the gauge degrees of freedom \eq{un}, 
we can always choose this gauge.} 
Thus the fluctuations $\delta A_\mu$ are essentially determined by $\delta
\Delta=\delta a$ up to gauge transformation. As discussed in \eq{gauge},
we have to require the gauge
condition \eq{gauge} for $\delta A_\mu (x)$. Let's
calculate the gauge condition \eq{gauge} 
for the solution \eq{varA} 
using \eq{normalization}, \eq{zero-mode}, \eq{invertible} and \eq{complete},
\bea \la{gauge-cal}
D_\mu \delta A_\mu &=& v^\dagger\p_\mu(v \delta A_\mu v^\dagger) v \xx
&=&-v^\dagger (\p_\mu \Delta f \Delta^\dagger \delta \Delta f \p_\mu
\Delta^\dagger-2\delta \Delta f \p_\mu \Delta^\dagger \Delta f 
\p_\mu \Delta^\dagger- \delta \Delta f \Delta^\dagger \p_\mu \Delta f 
\p_\mu \Delta^\dagger) v -{\rm h.c.}\xx
&=& -2 v^\dagger b f \Bigl({\rm tr}_2(\Delta^\dagger \delta \Delta) 
- {\rm tr}_2 (\delta \Delta^\dagger \Delta)\Bigr) f b^\dagger v =0,
\eea
where the following relations are used:
\bea \la{uf1}
&& {\bar \sigma}^\mu f \sigma^\mu=4f,\qquad \sigma^\mu b^\dagger
\Delta \sigma^\mu = -2 \Delta^\dagger b, \qquad {\rm for}\;
\Delta(x)=a - b {\bar {\bf x}},\\
\la{uf2}
&&  \sigma^\mu f {\bar \sigma}^\mu=4f,\qquad {\bar \sigma}^\mu b^\dagger
\Delta {\bar \sigma}^\mu = -2 \Delta^\dagger b, \qquad {\rm for}\;
\Delta(x)=a - b {\bf x}
\eea
and
\be \la{uf3}
{\bar \sigma}^\mu \Delta^\dagger \delta \Delta \sigma^\mu = 
2 {\rm tr}_2(\Delta^\dagger \delta \Delta),\qquad 
\sigma^\mu \delta \Delta^\dagger \Delta {\bar \sigma}^\mu = 
2 {\rm tr}_2(\delta \Delta^\dagger \Delta).
\ee
Thus the gauge condition \eq{gauge-cal} imposes another constraint 
\ct{mansfield} which we
denote as $\mu_g = {\rm tr}_2(\Delta^\dagger \delta \Delta) 
- {\rm tr}_2 (\delta \Delta^\dagger \Delta)=0$, or more explicitly,
\bea \la{gauge-adhm}
&&\mu_g= [\delta B_1,B_1^\dagger]-[B_1, \delta B_1^\dagger]+ 
 [\delta B_2, B_2^\dagger]-[B_2, \delta B_2^\dagger]\xx
&& \qquad \; +\delta II^\dagger - I \delta I^\dagger+ \delta J^\dagger J
-J^\dagger \delta J =0.
\eea
Note that the global $U(k)$ symmetry \eq{uk} is also fixed 
by the above gauge condition \eq{gauge-adhm} 
since it generates the ``global'' gauge transformation 
$a \rightarrow a + \delta a$.

Using the result \eq{varA}, one can find $\delta F_{\mu\nu}=D_\mu
\delta A_\nu - D_\nu \delta A_\mu$ as
\bea \la{varF}
\delta F_{\mu\nu}&=&v^\dagger \left\{ \p_\mu(v \delta A_\nu
  v^\dagger) - \p_\nu (v \delta A_\mu v^\dagger) \right\}v \xx
&=& v^\dagger \left\{ \p_\mu \Delta \delta f \p_\nu
  \Delta^\dagger - \p_\mu \Delta f \p_\nu \Delta^\dagger \Delta f \delta
  \Delta^\dagger- \delta \Delta f \Delta^\dagger \p_\mu \Delta f
  \p_\nu \Delta^\dagger \right\}v -(\mu \leftrightarrow \nu),
\eea
where $\delta f = -f(\delta \Delta^\dagger \Delta + \Delta^\dagger 
\delta \Delta) f$. 
To get the result \eq{varF}, we have used
\eq{normalization}, \eq{zero-mode}, \eq{invertible} and \eq{complete}
as in the derivation of \eq{gauge-cal}. 
One can easily check that the
resulting $\delta F_{\mu\nu}$ in \eq{varF} satisfies the
(anti-)self-dual equation \eq{zero-self} if one notices that the
function $\delta f(x)$ as well as $f(x)$ commutes with the quaternions 
since we required \eq{zero-adhm1} and \eq{zero-adhm2}, 
i.e. the factorization condition for $(\Delta + \delta \Delta)(x)$. 
So the fluctuations
$\delta A_\mu(x)$ in \eq{varA} describe the zero modes around a
classical instanton solution \eq{A} \ct{osborn-ap}. 
Explicitly, for example, for the anti-self-dual instanton with $\Delta(x)=a 
- b {\bar {\bf x}}$, the zero modes $\delta A_\mu$ has the following form
\be \la{exp-A}
\delta A_\mu =-v^\dagger(\delta a f \sigma_\mu b^\dagger - 
b {\bar \sigma}_\mu f \delta a^\dagger) v
\ee
and
\be \la{exp-F}
\delta F_{\mu\nu}=2iv^\dagger \left\{ b{\bar\sigma}^{\mu\nu} \delta f
  b^\dagger - b{\bar \sigma}^{\mu\nu} f b^\dagger \Delta f \delta
  \Delta^\dagger- \delta \Delta f \Delta^\dagger 
b{\bar \sigma}^{\mu\nu} f b^\dagger \right \}v.
\ee
Similarly, the expressions in the case of the self-dual background with 
$\Delta(x)=a - b {\bf x}$ can also be easily obtained.

The ADHM constraints \eq{zero-adhm1} and \eq{zero-adhm2} for zero modes
can be synthesized with \eq{gauge-adhm} into the constraints
\bea \la{zero-adhm3}
&& \delta \mu_k = [\delta B_1, B_1^\dagger]+ 
 [\delta B_2, B_2^\dagger] +\delta II^\dagger-J^\dagger \delta J =0, \\
\la{zero-adhm4}
&& \delta \mu_c = \delta IJ+ I\delta J
+[\delta B_1, B_2]+[B_1,\delta B_2]=0.
\eea
Since we now have the $4k^2$ constraints,
\eq{zero-adhm3} and \eq{zero-adhm4}, $\delta a$ has totally 
$4Nk$ real degrees of freedom, giving $4Nk$
zero modes as we already argued before. These are exactly given by
\eq{varA}. As an example, in the anti-self-dual background, 
they take the form \eq{exp-A}.

\section{Noncommutative Instantons and Their Fermionic Zero Modes}

Let's start with a brief review of commutative $SU(2)$ instantons. 
The standard ansatz for $A_\mu(x)$, for example the BPST \ct{bpst} or 
't Hooft ansatz \ct{thooft2}, 
entangles the group and Lorentz indices. This is basically
based on the fact that the Euclidean Lorentz group $O(4)$ is
isomorphic to $SU(2)_L \times SU(2)_R$, where two $SU(2)$ subgroups
correspond to the left-handed and right-handed chiral rotations.
More concretely, the Lorentz generators $L_{\mu\nu}=-L_{\nu\mu}$
can be decomposed into the self-dual
$(L^+_{\mu\nu}={1\over2}\varepsilon_{\mu\nu\rho\sigma}L^+_{\rho\sigma})$
and the anti-self-dual
$(L^-_{\mu\nu}=-{1\over2}\varepsilon_{\mu\nu\rho\sigma}L^-_{\rho\sigma})$
parts. According to the isomorphism $O(4) \cong SU(2)_L \times
SU(2)_R$, the Lorentz generators $L^+_{\mu\nu}$ and $L^-_{\mu\nu}$
can be mapped to $SU(2)_L$ and $SU(2)_R$, respectively. The
precise mapping can be achieved in terms of 't Hooft
$\eta$-symbol:
\begin{equation}\label{eta}
L^+_{\mu\nu} = \eta_{\mu\nu}^a T^a, \qquad L^-_{\mu\nu} = {\bar
\eta}_{\mu\nu}^a T^a,
\end{equation}
where $T^a$ are Lie algebra generators in $SU(2)_L$ or $SU(2)_R$. $T^a
={1\over2}\tau^a$ for spinors and $(T^a)_{bc}=-\varepsilon_{abc}$
for vector fields. It can be easily checked that each generator in
(\ref{eta}) separately satisfies the $O(4)$ algebra.

Under this decomposition the vector $A_\mu$ is indeed the
representation $({1\over2},{1\over2})$. Also the antisymmetric
tensor $F_{\mu\nu}$ in $O(4)$ having six components form a $(1,1)$
representation of $SU(2)_L \times SU(2)_R$. Then one can couple
the spinor indices of one of these $SU(2)_{L,R}$ subgroups to those of
$SU(2)_G$ \ct{jr}. (Here we use the subscript $G$ to mean the gauge group 
to distinguish another $SU(2)$'s in the Lorentz group.) 
Thus the (anti-)self-dual gauge field in $SU(2)_G$
gauge theory, i.e. $F^{\pm}_{\mu\nu}=\pm *F^{\pm}_{\mu\nu}$, may
be obtained by reshuffling several $SU(2)$'s, i.e. $SU(2)_{L,R}$ and
$SU(2)_G$:
\begin{equation}\label{F}
F^{+a}_{\mu\nu}(x)=\eta_{\mu\nu}^a f(x^2),\qquad
F^{-a}_{\mu\nu}(x)={\bar \eta}_{\mu\nu}^a f(x^2),
\end{equation}
where $f(x^2)$ is a scalar function satisfying some harmonic equation.
Thus group theory essentially determines the tensor structure of
(anti-)self-dual gauge fields. (Actually this is also a reason 
why quaternions $(\cong SU(2))$ play a crucial role 
in the ADHM construction.)

When we try to generalize the above consideration to the
noncommutative space \eq{NC-space}, what happens ?  
Since $\theta^{\mu\nu}$ are anti-symmetric tensor, let's
decompose them into self-dual and anti-self-dual parts:
\begin{equation}\label{theta}
\theta_{\mu\nu} = \eta_{\mu\nu}^a \zeta^a + {\bar \eta}_{\mu\nu}^a
\chi^a.
\end{equation}
Note that, as seen from \eq{adhm1}, the ADHM construction depends
only on the self-dual or the anti-self-dual part of $\theta_{\mu\nu}$ 
because $\eta^a_{\mu\nu}{\bar \eta}^b_{\mu\nu}=0$. 
Since the self-duality condition is invariant under $SO(4)$ rotations
(or more generally $SL(4,{\bf R})$ transformations),
one can always make the matrix $\theta_{\mu\nu}$ to a standard
symplectic form by performing the $SO(4)$ transformation $R$: 
\be
\theta=R {\tilde \theta} R^T,
\ee
where we choose ${\tilde \theta}$ as
\be \la{theta}
{\tilde \theta}_{\mu\nu} = \left(
\begin{array}{cccc} 
0 & \theta_1 & 0 & 0 \\
-\theta_1 & 0 & 0 & 0 \\
0 & 0 & 0 & \theta_2 \\
0 & 0 & -\theta_2 & 0 
\end{array}
\right).
\ee
There are four important cases to consider:
\bea \la{c-space}
&& \bullet \; \theta_1=\theta_2=0: \quad \mbox{commutative ${\bf R}^4$},\\
\la{nc-sd}
&& \bullet \;  \theta_1=\theta_2={\zeta \over 4} : 
\quad \mbox{self-dual $\do$},\\
\la{nc-asd}
&& \bullet \;  \theta_1=-\theta_2={\zeta \over 4}: 
\quad \mbox{anti-self-dual $\do$},\\
\la{nc-22}
&& \bullet \;  \theta_1\theta_2=0 \; \mbox{but}\;  \theta_1+\theta_2 \neq 0: 
\quad \re \times \mi.
\eea

By the noncommutativity \eq{NC-space}, the original Lorentz symmetry
is broken down to its subgroup. What are the remaining spacetime symmetries 
for each cases ? \footnote{Note that, by the canonical choice
  \eq{theta} of the noncommutativity, we already fixed our coordinate
  system. So we are considering symmetry properties with respect to this
coordinate system.} 
We will refer only rotational symmetry since the noncommutative space
\eq{NC-space} always respects a global translational symmetry.
For the commutative ${\bf R}^4$, of course, we have full $O(4)$
symmetry. However, for the self-dual and the anti-self-dual $\do$, 
the original Lorentz symmetry $O(4) \cong SU(2)_L \times SU(2)_R$ is
broken down to the subgroup:
\begin{equation}\label{breaking1}
  SU(2)_L \times SU(2)_R \rightarrow 
\left \{ \begin{array}{l} SU(2)_R \times U(1)_L,
\qquad \mbox{self-dual $\do$},\\
  SU(2)_L \times U(1)_R, \qquad \mbox{anti-self-dual $\do$}.
\end{array} \right.
\end{equation}
On the other hand, $\re \times \mi$ has more complicated Lorentz
symmetry breaking:
\be \la{breaking2}
SU(2)_L \times SU(2)_R \rightarrow 
\left\{ \begin{array}{l}
SU(2)^{-}_D \times U(1)_D^{+}, \qquad \mbox{for $\theta_1 \neq 0$},\\
SU(2)^{+}_D \times U(1)_D^{-}, \qquad \mbox{for $\theta_2 \neq 0$},
\end{array} \right.
\ee
where $SU(2)^{\pm}_D$ are the diagonal subgroups of $O(4)$ generated by
$L^+_{\mu\nu} \pm L^-_{\mu\nu}$ and $U(1)_D^\pm$ are their $U(1)$
subgroups.

Let's specify $U(N)$ instantons in the noncommutative 
space \eq{theta}. From \eq{adhm1}, one sees that anti-self-dual
(self-dual) instantons on self-dual (anti-self-dual) $\do$ are
described by deformed ADHM equation and the singularities of
instanton moduli space $\CM_{k,N}$ are resolved \ct{ns,sw}. 
While self-dual
(anti-self-dual) instantons on self-dual (anti-self-dual) $\do$ are
described by undeformed ADHM equation and the singularities of
instanton moduli space $\CM_{k,N}$ still remain. While the
former is not a BPS state, the latter is a BPS state.  
For the $\re \times \mi$ space discussed in \ct{kly2,ckt}, 
both self-dual and anti-self-dual $U(N)$ instantons are
described by the deformed ADHM equation and their moduli
space is always non-singular. This case is also a non-BPS state.

For $U(1)$ gauge theory, according to a stability theorem 
by Nakajima \ct{naka}, 
we can always put $J=0$. So the ADHM contraints \eq{adhm3} and
\eq{adhm4} in the $U(1)$ case reduce to in the basis \eq{theta}
\bea \la{u1-adhm1}
&& \mu_r=[B_1,B_1^\dagger]+[B_2,B_2^\dagger]+II^\dagger
=\theta^{\mu\nu}\eta^3_{\mu\nu} \\
\la{u1-adhm2}
&& \mu_c=[B_1,B_2]=0.
\eea
The most general solution for \eq{u1-adhm1} and \eq{u1-adhm2} 
may be obtained by simultaneously triangularizing the matrices
$B_1$ and $B_2$ \footnote{We are grateful to Kimyeong Lee 
for pointing out a mistake in the earlier version of this paper.} 
while they can be simultaneously diagonalized 
when all instanton separations become large, i.e.
\be \la{diabb}
B_1={\rm diag}(z_1^1, \cdots, z_1^k),\qquad 
B_2={\rm diag}(z_2^1, \cdots, z_2^k).
\ee
A special solution is that one of them is identically zero, e.g. $B_2=0$.
This solution describes $k$ instantons which sit along
the complex $z_1$-plane, so-called elongated instantons \ct{bn,kuro}. 
Note that the moduli space $\CM_{k,1}$ for $k\;U(1)$ instantons is
completely determined by translations of instantons, that is 
$4k$ position moduli.

Depending on the value of $\theta^{\mu\nu}$, $U(1)$ instantons have
different properties as discussed above. 
If $\do$ is self-dual, then $\mu_r=\zeta$. This case describes 
the Nekrasov-Schwarz instanton with fixed size $\sqrt{\zeta}$ \ct{ns}. 
On the other hand, in the case of anti-self-dual $\do$, $\mu_r=0$ and
$I=0$, which describes localized instantons with zero size 
\ct{agms,hkl,kf3,haoo,ham}. 
Since the structure of commutative instantons has been almost
determined by symmetry property itself, the immediate question
is whether one can understand noncommutative instantons, especially
$U(1)$ instantons, by the symmetry breaking
(\ref{breaking1}) or (\ref{breaking2}). 
We think this may be the case although we don't
arrive at concrete answer yet. But, let's just quote 
an observation that at least the anti-self-dual (or the self-dual) 
single $U(1)$ instanton has this $SU(2)$ algebra structure 
in the self-dual (or in the anti-self-dual) $\do$ (see also
\ct{gms}). For example, for the anti-self-dual $U(1)$ instanton
in the self-dual $\do$, with the notation (22) in \ct{kly1} 
(where $\zeta=2$),
\begin{equation}\label{U1}
  F_A=-\frac{2i}{x^2(x^2+1)(x^2+2)}{\bar \eta}_{\mu\nu}^a T^a
  dx^\mu \wedge dx^\nu,
\end{equation}
where $T^+=-T^1 + iT^2=z_1\bz_2, \;T^-=-T^1 - iT^2=\bz_1z_2$ and 
$T^3= \half (z_1 \bz_1-z_2\bz_2)$ and thus $[T^a,
T^b]=-i\varepsilon^{abc} T^c$. \footnote{This $SU(2)$ algebra may be 
viewed as a noncommutative Hopf fibration, as in \ct{ccly}, 
$\pi: {\bf S^3} \cong SU(2)_L \rightarrow {\bf S}^2 
\cong SU(2)_L/U(1)_L$ related to the symmetry 
breaking \eq{breaking1}.} 
Since $T^a$'s commute with $x^2$, 
the solution \eq{U1} really looks like a commutative $SU(2)$
instanton where the gauge group is given by the unbroken $SU(2)_R$
spacetime symmetry. Besides, the (anti-)self-dual $U(1)$ localized 
instantons are purely determined by the unbroken $U(1)_{L,R}$
symmetry.

As we discussed in Section 2, the whole family of the solutions of 
\eq{zero-adhm3} and \eq{zero-adhm4} gives $4Nk$ moduli. 
It is difficult to completely solve the constraints
\eq{zero-adhm3} and \eq{zero-adhm4} to find the zero modes 
of $U(1)$ instantons. However, when all distances between 
$k \; U(1)$ instantons (both BPS and non-BPS instantons) become large, 
$\delta B_{1,2}={\rm diag}
(\delta z_{1,2}^1,\cdots,\delta z_{1,2}^k)$
and $\delta I =0$ solve the constraints 
\eq{zero-adhm3} and \eq{zero-adhm4} since $B_1$ and $B_2$ can be 
diagonalized in the limit as in \eq{diabb}. The zero modes 
in the dilute $k$
anti-self-dual instantons are so given by \eq{exp-A} with $4k$
eigenvalues of $\xi_\mu$ and $\delta \lambda =\delta \lambda^\dagger =0$. 
For the similar reason, it is difficult to solve 
the ADHM constraints \eq{zero-adhm3} and \eq{zero-adhm4} completely 
for $k \; U(N)$ instantons for which $\delta B_{1,2}$ can be
diagonalized only for dilute instanton gas limit.

As observed in \ct{kf1,kly1}, 
a $U(N)$ instanton always contains a $U(1)$ instanton, 
the Nekrasov-Schwarz instanton or the localized instanton, 
in the limit that the size of $SU(N)$ instanton vanishes 
which corresponds to the limit $J \to 0$. \footnote{For example, in
  the case of single $U(2)$ instanton, the ADHM constraints \eq{adhm3}
  and \eq{adhm4} can be solved by $I=\sqrt{\rho^2+\zeta} 
(\cos\alpha e^{i\beta},\sin\alpha e^{-i\beta}),\; 
J^\dagger =\rho(\sin\alpha e^{i(\beta-\gamma)},
-\cos\alpha e^{-i(\beta+\gamma)})$ where $\rho$ is the size of an
$SU(2)$ instanton and $(\alpha, \beta, \gamma)$ are global $SU(2)$
angles. Here we fixed the global $U(1)$ symmetry in \eq{uk}.}
In this limit, the matrices $B_1$ and $B_2$ can be
diagonalized like as \eq{diabb} if all instantons are 
sufficiently distant. Then we see that 
$\delta B_1$ and $\delta B_2$ should also be diagonalized 
to satisfy \eq{zero-adhm4} since they are independent of each other.
This requires that $\delta I=0$ because of \eq{zero-adhm3}. So, in the
limit $J \to 0$ and in the dilute instanton gas limit, 
the moduli from $\lambda$ have to be frozen and
only $4k$ moduli from $B_1$ and $B_2$ remain. Eventually we thus
arrive at the moduli space of $U(1)$ instantons. At this stage, an
intriguing question arises. Since $\CM_{k,N}$ for $k$ noncommutative 
instantons is $4Nk$-dimensional, this implies that the size 
$\rho$ of an $SU(N)$ instanton should be one of the $4N$ moduli.
According to \ct{jnr}, we can guess that the size moduli 
of noncommutative instantons should be generated by conformal
transformation. This naturally implies that still there exist
corresponding (super)conformal zero modes in the noncommutative 
instanton background. However we will show that 
the (super)conformal zero modes in non-BPS 
instantons are affected by the noncommutativity.

It is easy to find some special zero modes satisfying 
\eq{zero-adhm3} and \eq{zero-adhm4} which are related to 
supersymmetric fermionic zero modes \ct{jr,nsvvz}. 
They are given by 
\be \la{zero-tr}
\delta a = b {\bar \sigma}_\mu c_\mu
\ee
where $c_\mu$ is a constant real four-vector. 
In order to check \eq{zero-adhm3} and \eq{zero-adhm4} for
\eq{zero-tr}, it is more convenient to directly check 
using \eq{sigma} that 
$\delta \mu^a = {\rm tr}_2 \tau^a (\delta a^\dagger a 
+ a^\dagger \delta a)=0$ and $\mu_g = {\rm tr}_2(\Delta^\dagger \delta a
- \delta a^\dagger \Delta)=0$. The solution \eq{zero-tr} gives 
$\delta A_\mu= F_{\mu\nu}^{cl} c_\nu$ in \eq{exp-A} 
which have the anti-self-dual 
field strength $\delta F_{\mu\nu}
=-D_\lambda F_{\mu\nu}^{cl} c_\lambda$ 
proved using the Bianchi identity and 
$D_\mu \delta A_\mu= D_\mu F_{\mu\nu}^{cl} c_\nu =0$ 
by the equations of motion.

In the case of commutative instantons, we can have additional
zero modes which are related to ``superconformal zero-modes''  \ct{jr,nsvvz}:
\be \la{zero-cf}
\delta a = a {\bar \sigma}_\mu d_\mu + {b \over k}{\rm tr}_k(b^\dagger a)
\ee
where ${\rm tr}_k$ is the trace over $k \times k$ matrices and 
$d_\mu$ is a constant real four-vector.
However, one can check that, in the case of non-BPS instantons, 
\eq{zero-cf} doesn't satisfy \eq{zero-adhm3} and
\eq{zero-adhm4} and no longer gives zero modes. 
To see this, let's calculate 
$\delta \mu^a$ and $\mu_g$ for the ansatz \eq{zero-cf}. It is sufficient to
check the first term in \eq{zero-cf} since the second term 
has the same structure with \eq{zero-tr}.
A simple calculation gives
\bea \la{nczero-cf}
&& \delta \mu^a = 2 \theta^{\mu\nu}(\eta_{\mu\nu}^a d_4 
+ \varepsilon^{abc} \eta_{\mu\nu}^b d_c) \xx
&& \mu_g =-2i \theta^{\mu\nu}\eta^a_{\mu\nu} d_a.
\eea
There is no solution satisfying $\delta \mu^a=0$ and $\mu_g=0$ 
if $\theta^{\mu\nu}$ is self-dual or $\re \times \mi$. 
The same conclusion arises in the self-dual instanton.
Thus, for the reason discussed in Section 2, the fluctuation
\eq{zero-cf} in the non-BPS background fails to give a field strength
satisfying \eq{zero-self}. \footnote{One may explicitly verify 
using $a^\dagger a = (f^{-1} - x^2 + 2\xi_\mu x_\mu) \otimes {\bf
    1}_2 + \half \theta^{\mu\nu} \sigma_{\mu\nu}$ 
that the first term in \eq{varF} breaks the self-duality property.}  
Of course, for BPS instantons, \eq{zero-cf} still gives zero modes because 
${\bar \eta}_{\mu\nu}^a \eta_{\mu\nu}^b =0$.
It could be originated from the fact that the non-BPS instanton has a 
minimum size set by the noncommutative scale.
Anyway it remains an open problem how to modify the ansatz 
\eq{zero-cf} to satisfy $\delta \mu^a=0$ and $\mu_g=0$ for all cases.
\footnote{One may try to add a term to cancel the right-hand side of 
\eq{nczero-cf}, for example, 
such as $-\half a (a^\dagger a)^{-1} \theta^{\mu\nu}
\sigma_{\mu\nu} {\bar \sigma}_\lambda d_\lambda.$ Note that $a^\dagger
a$ is invertible since $\Delta^\dagger \Delta$ is so in whole space, e.g. 
${\bf x}=0$. One may find
that now $\delta \mu^a =0$ but $\mu_g \neq 0$.
If one simply chooses $\delta a = a-\half a (a^\dagger a)^{-1} \theta^{\mu\nu}
\sigma_{\mu\nu}$ instead,  
one will also get $\delta \mu^a = 0,\; \mu_g \neq 0$.}

Now let's find the normalizable fermionic zero modes by
solving a massless Dirac equation in the background of anti-self-dual 
$k$ instantons.
Following the same strategy in \ct{bcl}, \eq{gauge} and \eq{zero-self}
can be united into
\be \la{dirac-ad}
D_\mu {\rm tr}_2(\sigma^\mu \Phi_1 \sigma^\nu) =0
\ee
where the matrix $\Phi_1$ has the form
\be \la{Phi1}
\Phi_1 ={\bar \sigma}^\mu \delta A_\mu = \left (\begin{array}{cc}
\delta A_4 -i\delta A_3 & -\delta A_2 - i\delta A_1 \\
\delta A_2 -i \delta A_1 &  \delta A_4 + i\delta A_3
\end{array} \right)
\ee
and thus $\Phi_1$ satisfies the reality condition
\be \la{real}
\sigma^2 \Phi^*_1 \sigma^2 =- \Phi_1.
\ee
Since the $\sigma^\nu$ are a complete set of $2 \times 2$ matrices,
\eq{dirac-ad} can be reduced to
\be \la{dirac-ad1}
\sigma^\mu D_\mu \Phi_1 =0.
\ee
Since the field equation \eq{dirac-ad1} independently acts 
on each column of $\Phi_1$ and the second column is uniquely 
determined by the reality condition \eq{real} from the first column,
one can solve the field equation \eq{dirac-ad1} in terms of a
two-component spinor 
\be \la{chi}
\chi = \left( \begin{array}{c}
\delta A_4 -i\delta A_3 \\
\delta A_2 -i \delta A_1
\end{array} \right)
\ee
that obeys
\be \la{dirac-ad2}
\sigma^\mu D_\mu \chi =0.
\ee
Thus the spinor $- \sigma^2 \chi^*$ which is the second column of
$\Phi_1$ is also a solution of the field equation \eq{dirac-ad2}.
From the solution $\chi$, one can construct a second, linearly
independent matrix solution $\Phi_2$ obtained by replacing $\chi$ by
$i\chi$:
\be \la{Phi2}
\Phi_2 =\Phi_1 \sigma^3 = \left (\begin{array}{cc}
\delta A_3 + i\delta A_4 & -\delta A_1 + i\delta A_2 \\
\delta A_1 +i \delta A_2 &  \delta A_3 - i\delta A_4
\end{array} \right).
\ee 
In sum, for each linearly independent spinor solution, there are
precisely two linearly independent solutions about the fluctuation
$\delta A_\mu$. So the problem of counting the number of bosonic
zero modes is now translated to that of counting the fermionic
zero modes in \eq{dirac-ad2}, or calculating the index of the Dirac
operator.

As counted in Section 2, there are totally $4Nk$ bosonic zero modes in
the $k$ instanton fields. Thus we should have $2Nk$ adjoint fermion
zero modes satisfying \eq{dirac-ad2}.
We have already known from \eq{varA} and \eq{varF} 
what $\delta A_\mu$ in \eq{chi} are and that they satisfy the
self-duality condition \eq{zero-self}. Thus if we plug the solution 
\eq{varA} in \eq{chi}, then the resulting spinor $\chi$ automatically
satisfies the Dirac equation \eq{dirac-ad1}. Thus we can easily construct
$2Nk$ adjoint fermion zero modes in this way.

The zero modes \eq{zero-tr} can be translated into an adjoint
fermion zero mode
\begin{equation} \la{fzero-tr}
\chi = {\bar \sigma}^\mu \delta A_\mu \pmatrix{1 \cr 0 }
=\frac{i}{2} F_{\mu\nu}^{cl} {\bar \sigma}^{\mu\nu}\epsilon.
\end{equation}
Here we expressed the translation parameters $c_\mu$ in terms of the
two-component spinor $\epsilon$: 
$c_\mu := \sigma_\mu \epsilon$. Actually $\epsilon$ plays the role of
supersymmetric partner of instanton position moduli \ct{nsvvz}. Since the
translational symmetry is still manifest in noncommutative space
\eq{NC-space} and the supersymmetry is at most softly broken 
by the noncommutativity, 
the position moduli must have the supersymmetric partner.  
However, we observed that for the non-BPS
instanton a supersymmetric partner corresponding to the size of
instanton is affected by the noncommutativity.
Since the non-BPS instanton has a minimum size set by the
noncommutativity and this is just that of a $U(1)$ instanton, 
the superconformal symmetry as well as the conformal symmetry has to act 
nontrivially only on the space of $SU(N)$ instantons. 
But in noncommutative space it is not
possible to separate $U(N)$ into $SU(N)$ and $U(1)$. 
So the (super)conformal transformation in the noncommutative 
instanton background may 
be a challenging problem. The issue on the instanton calculus 
in noncommutative super Yang-Mills
theory, especially the zero modes related to the superconformal symmetry 
as well as the conformal symmetry, definitely deserves further study.

If $q$ denotes the fundamental representation of $U(N)$, the adjoint
representation can be obtained by decomposing $q \otimes {\bar q}$,
for which
\be \la{coder-2f}
D_\mu =\p_\mu + A_\mu \otimes 1 + 1 \otimes {\bar A}_\mu.
\ee
Following the same strategy in \ct{cgt} to get instantons 
in higher dimensional representation,\footnote{It is quite 
plausible that the tensor product of instantons {\it \'a la} 
\ct{cgt} is still applicable to noncommutative instantons 
since the ADHM construction is complete even for noncommutative 
instantons; the (anti-)self-dual solution for $G_1 \times G_2$ 
may be considered as a particular solution for a larger group $G$ 
containing $G_1 \times G_2$.
This speculation may be understood more clearly by considering
the brane configuration of corresponding $D4/D0$ system 
in the 2-form $B_{\mu\nu}^{NS}$ background. 
Even though explicit demonstration on 
this is an interesting problem, however, 
it goes beyond the scope of this paper.} one may show that the 
zero modes \eq{dirac-ad2} for the adjoint fermion can be constructed by
the zero modes in the fundamental representation according to the
correspondence \eq{coder-2f}. So it will be interesting to consider 
the Dirac equation for a fermion $\eta$ in the fundamental 
representation with positive chirality 
\be \la{fchiral-dirac}
\sigma^\mu D_\mu \eta =\sigma^\mu(\p_\mu + v^\dagger \p_\mu v) \eta=0.
\ee
Transposing \eq{fchiral-dirac} with respect to spinor
indices and then multiplying it by $v$ on the left and $\sigma^2$
on the right, we can rewrite \eq{fchiral-dirac} as 
\be \la{fchiral}
P\p_\mu ({\tilde \eta}{\bar \sigma^\mu})=0
\ee
where $P=v v^\dagger$ and
\be \la{eta}
{\tilde \eta}=v \eta^T \sigma^2.
\ee
Using \eq{zero-mode} and \eq{complete}, one can show that \footnote{The
  proof of \eq{fadhm} is exactly the same as the commutative case
  \ct{cfgt,osborn,csw,osborn-ap}. More generally, if we introduce $k$
  (complex) vectors in the fundamental representation 
of $U(N)$, $a_\mu = v^\dagger b f 
{\bar \sigma}_\mu$, then $D_\mu a_\nu =v^\dagger \partial_\mu
(Pbf {\bar \sigma}_\nu) = v^\dagger bf \xi_\lambda f 
(2 \delta_{\lambda\nu}{\bar\sigma}_\mu +
2 \delta_{\lambda\mu}{\bar\sigma}_\nu- {\bar\sigma}_\mu \sigma_\nu
{\bar \sigma}_\lambda)$, so that $f_{\mu\nu} = 
-*f_{\mu\nu}, \; D_\mu a_\mu=0$ where $f_{\mu\nu}=
D_{\mu}a_{\nu}-D_\nu a_\mu$. Thus $a_\mu$ gives $2k$ (real) vector
zero modes in the fundamental representation \ct{osborn-ap}.} 
\be \la{fadhm}
P\p_\mu(Pb{\bar \sigma}_\mu f)=0.
\ee
Hence uniqueness leads us to $k$ independent solutions for $\eta^T$ 
as an $N \times 2$ matrix
\be \la{fzero-sol}
\eta^i_{u,\alpha}=(v^\dagger b f \sigma^2)_{u,i\alpha}
\ee
where $u=1,\cdots,N, \; i=1, \cdots, k$ and
$\alpha=1,2$. \footnote{One can check that the zero modes \eq{fzero-sol} 
for $U(1)$ case exactly reproduce the solution (4.5) in \ct{nekra} if
one notices that our $v$ corresponds to $\Psi$ in \ct{nekra}.} 
Thus we have found $k$ fermionic zero modes 
in the fundamental representation.

\section{Atiyah-Singer Index in Noncommutative Instantons}

If we introduce the Euclidean Dirac matrices \eq{gamma}, the spinor
equation \eq{dirac-ad2} or \eq{fchiral-dirac} is equivalent 
to the Dirac equation
\be \la{dirac-eq}
\gamma^\mu D_\mu \psi_+=0
\ee
for a spinor field $\psi_+$ with definite chirality, 
\be \la{chiral}
\gamma_5 \psi_+ = \psi_+
\ee
and transforming in the adjoint or fundamental representation of
$U(N)$, respectively. Here $D_\mu$ is the covariant derivative in
the instanton background. In the previous section we showed that 
in noncommutative $k$ instanton background in $U(N)$ gauge theory the
number of fermion zero modes in the adjoint representation and the
fundamental representation are respectively $2Nk$ and $k$.
In this section we will show that the number of the fermionic 
zero modes given by (\ref{dirac-eq}) is related to the topological 
charge of the instanton gauge field by the so-called Atiyah-Singer index
theorem \ct{atiyah}. A reinterpretation of this theorem within the instanton
calculus \cite{bcl} shows that this theorem is perfectly
equivalent to the anomaly in axial-vector current related to a
regularization of the triangular graph. The interrelation between
the Atiyah-Singer index, the chiral anomaly and the zero modes of Dirac
operator in topologically nontrivial gauge fields form a Golden
Triangle. Our calculation here is essentially the same as the
chiral anomaly calculation in \ct{as,gbm}.

We start with the Dirac equation \eq{dirac-eq} for a Dirac 
fermion instead in
an arbitrary representation (fundamental, adjoint, etc) in the
background of anti-self-dual instantons
\be \la{psi}
\Psi= \pmatrix{\psi_+ \cr \psi_-}.
\ee
In the basis \eq{gamma} 
\be \la{chiral}
\gamma_5 \psi_{\pm}=\pm \psi_{\pm}.
\ee
Since
\be \la{DD}
(\gamma D)^2= D^2 +\frac{i}{2}\left(\begin{array}{cc}
{\bar \sigma}^{\mu\nu}F_{\mu\nu}^{cl} & 0 \\
0 & \sigma^{\mu\nu}F_{\mu\nu}^{cl} \end{array}\right),
\ee
in the anti-self-dual instanton background, i.e. 
$\sigma^{\mu\nu}F_{\mu\nu}^{cl}=0$, only the positive chirality spinor 
$\psi_+$ can be zero modes. \footnote{In fact, the zero modes $\psi_+$
  will be given by $\chi$ in \eq{chi} in the adjoint representation, 
so $2Nk$ solutions, and by $\eta$ in \eq{fzero-sol} in the fundamental
representation, so $k$ solutions. The space consisted of these modes 
is exactly ${\rm ker} D\!\!\!\!/$ and thus they contribute to the
index of the Dirac operator as we will discuss.} 
This means that 
${\rm ker} D\!\!\!\!/ ={\rm ker} {\bar D\!\!\!\!/} \, D\!\!\!\!/ \;$ 
where $D\!\!\!\!/ = \sigma^\mu D_\mu,\; {\bar D\!\!\!\!/}=  
{\bar \sigma}^\mu D_\mu$ since ${\rm ker} {\bar D\!\!\!\!/} =0$.

We can count the number of zero modes using index theorem. The index
of the Dirac operator is defined as
\be \la{index}
{\rm Ind} D\!\!\!\!/ = {\rm dim}\{{\rm ker} D\!\!\!\!/ \} 
-  {\rm dim}\{{\rm ker} {\bar D\!\!\!\!/} \}.
\ee
In order to calculate the index \eq{index}, 
let's consider the quantity \ct{bcl,jr}
\be \la{anomaly} 
T(m^2)=\Tr \frac{m^2}{-(\gamma D)^2 + m^2} \gamma_5 = 
\Tr \Biggl \{ \frac{m^2}{-{\bar D\!\!\!\!/} \, D\!\!\!\!/ + m^2}
-\frac{m^2}{- D\!\!\!\!/ \, {\bar D\!\!\!\!/} + m^2} \Biggr\}
\ee where the trace denotes a diagonal sum over all indices,
e.g. group indices, spinor indices and 
$\Tr_\CH$, an integration over spacetime. One can see that $T(m^2)$ is
independent of $m^2$ and indeed exactly equal to the index 
defined in \eq{index}. The reason is following \ct{bnv}. If $\psi$ is an
eigenfunction of ${\bar D\!\!\!\!/} \, D\!\!\!\!/$, then $D\!\!\!\!/
\, \psi$ is an eigenfunction of $D\!\!\!\!/ \, {\bar D\!\!\!\!/}$
with the same eigenvalue. Conversely, if $\psi$ is an
eigenfunction of $D\!\!\!\!/ \, {\bar D\!\!\!\!/}$, 
then ${\bar D\!\!\!\!/} \, \psi$ is an eigenfunction of 
${\bar D\!\!\!\!/} \, D\!\!\!\!/ \,$ with the same eigenvalue. 
This means that there is a pairwise
cancellation in \eq{anomaly} coming from the sum over eigenstates
with non-zero eigenvalues. 
So the only contribution is coming from the zero modes, for which the
first term simply gives one for each zero mode and the second term
vanishes because ${\rm ker} {\bar D\!\!\!\!/} =0$. 
Thus the result is clearly
independent of $m^2$ and moreover it is exactly equal to 
${\rm Ind} D\!\!\!\!/
\;$ since ${\rm ker} {\bar D\!\!\!\!/} \, D\!\!\!\!/ 
= {\rm ker} D\!\!\!\!/$.

Since $T(m^2)$ is independent of $m^2$, we can evaluate it in two different
limits, i.e. $m^2 \to 0$ and $m^2 \to \infty$,
\bea \la{AST} 
{\rm Ind} D\!\!\!\!/ &=& \lim_{m^2 \to 0} T(m^2)=\Tr P_0 \gamma_5 \xx 
&=& \lim_{m^2 \to \infty} T(m^2)=\frac{1}{16\pi^2} \int d^4 x
F_{\mu\nu}^{cl\,a} *F_{\mu\nu}^{cl\,b} {\rm tr}(T^aT^b), 
\eea 
where $P_0$ is a projection operator into the subspace of zero modes.
Here the calculation of $T(m^2)$ in the large $m^2$ limit is exactly
the same as the commutative case. In this calculation 
we don't meet any trouble due to the
noncommutativity. So we will not repeat it here but, for example, 
see \ct{bcl}. In the fundamental representation we normalize the
generators $T^a$ of $U(N)$ so that ${\rm tr}(T^aT^b) =-\half \delta^{ab}$.
Other representations are constructed by decomposing tensor product of
these and, in particular, for the adjoint representation of $U(N)$,
${\rm tr}(T^aT^b) =- N \delta^{ab}$. So we get 
\be \la{as}
{\rm Ind} D\!\!\!\!/ =n_+ - n_-= 
\left \{ \begin{array}{l} -2Nk,
\qquad \mbox{adjoint} \\
- k,\qquad \mbox{fundamental}
\end{array} \right. \\
\ee
where $n_+(n_-)$ is the number of zero modes 
with positive (negative) chirality. In the anti-self-dual background,
$n_-=0$ while $n_+=0$ in the self-dual background as we discussed above.

\section{Discussion}

In this paper we discussed bosonic and
fermionic zero modes in noncommutative instanton backgrounds 
based on the ADHM construction. We showed that 
the number of instanton zero modes is precisely equal to 
the Atiyah-Singer index of chiral Dirac operator. 
We pointed out that (super)conformal
zero modes in the non-BPS background are affected by the
noncommutativity and so their explicit construction in noncommutative
space remains an open problem. We would like to discuss this problem a
little more.

As shown in this paper, the moduli space $\CM_{k,N}$ of 
$k$ instantons in $U(N)$
gauge theory is $4kN$-dimensional. If we specialize to single $U(2)$
instanton, $\CM_{1,2}$ is 8-dimensional. By the comparison to
commutative $SU(2)$ instanton, we know that the size parameter 
is still a modulus of the noncommutative instanton.
Thus we can guess that it will be generated by the conformal
transformation, more precisely, by dilatation, as in the commutative
case \ct{jnr}. 
However, as we discussed in Section 4, the size modulus should be
frozen in the $U(1)$ instanton limit which is defined by $J \to 0$ in
this paper. This implies that the scale
transformation has to act nontrivially only on the $SU(2)$ instanton sector. 
In other words, the dilatation acting on the $U(1)$ instanton 
probably generates only pure gauge transformation and thus it can be
gauged away. However, because only $U(N)$ algebra is closed 
in noncommutative space and so it is not possible 
to separate $U(1)$ and $SU(N)$, it is not obvious how to 
define the conformal transformation for the instanton moduli.
The structure of Lorentz symmetry breaking in \eq{breaking1} and
\eq{breaking2} may be helpful to shed light on this problem.  
Anyway we put this interesting problem for future work.

We speculated that $U(1)$ instantons may be understood 
by the structure of Lorentz symmetry breaking in \eq{breaking1} and
\eq{breaking2}. Let's discuss it a little more. 
For specification, let's consider the self-dual $\do$ in which 
$\theta_{\mu\nu}=\eta_{\mu\nu}^a \zeta^a$.  In this case 
we have the following Lorentz symmetry breaking:
\be \la{lsb}
SU(2)_L \times SU(2)_R \rightarrow 
SU(2)_R \times U(1)_L.
\ee
We observed in \eq{U1} that the unbroken $SU(2)_R$ plays 
a role of gauge group. 
Note that the size of non-BPS $U(1)$ instanton is given by 
$(\frac{1}{4} \theta^{\mu\nu}\theta_{\mu\nu})^{\frac{1}{4}}
=(\zeta^a\zeta^a)^{\frac{1}{4}}:=\sqrt{\zeta}$.
From \eq{lsb}, we can imagine $\zeta^a:=\vec{\zeta}$ is a vector 
in the $SU(2)_L$ space.
Since we know that the magnitude of $\vec{\zeta}$ is constant, the
allowed values of $\vec{\zeta}$ should lie on ${\bf S}^2$ with radius $\zeta$. 
Regardless of what direction $\vec{\zeta}$ is pointing in $SU(2)_L$
space, it is invariant with respect to $U(1)_L$ rotations about the
$\hat{\vec{\zeta}}$-axis. Thus the allowed values of $\vec{\zeta}$ are
parameterized by $SU(2)_L/U(1)_L \cong {\bf S}^2$ 
(compare with the footnote 8). This was the reason why the relative metric
of two $U(1)$ instantons is the Eguchi-Hanson \ct{lty}. In this
spirit, the $U(1)$ instanton here is very similar to the Abelian instanton
obtained by a bundle reduction due to symmetry breaking for Einstein
manifolds, as discussed by Soo in \ct{soo}. Also Braden and Nekrasov 
\ct{bn} argued that the noncommutative $U(1)$ instanton corresponds 
to a non-singular $U(1)$ gauge field on a commutative K\"ahler
manifold obtained by a blowup of ${\bf C}^2$ at a finite number of
points. It will be interesting to study the noncommutative $U(1)$
instantons from these perspectives.

In order to calculate instanton effects in quantum gauge theory, it is
important to know the Green function in instanton backgrounds. 
In a pioneering work \ct{bccl}, Brown, Carlitz, Creamer, and Lee showed
that the propagators for massless spinor and vector fields are
determined by the massless scalar propagators. And the scalar
propagator $G(x,y)$ has a remarkably simple expression
\be \la{green}
G(x,y)=v(x)^\dagger G^{(0)}(x,y) v(y)
\ee
where $G^{(0)}(x,y)=\frac{1}{4\pi^2(x-y)^2}$ is the Green function 
for the ordinary Laplacian, i.e. $-\p_\mu \p_\mu
G^{(0)}(x,y)=\delta(x-y)$. It is not difficult to prove 
\ct{cfgt,osborn,csw} that  
$-D_\mu D_\mu G(x,y)=\delta(x-y)$. An immediate question is how to
generalize the Green function \eq{green} 
to noncommutative instanton backgrounds. 
The free scalar Green function
was already defined in noncommutative space in \ct{chms,gn}. 
Using this Green function, the
propagators in the noncommutative instanton background may be obtained
as long as an annoying ordering problem is carefully treated. 
(Recently we solved this problem in \ct{ly}.) 
We hope to address these problems mentioned here in
near future.


\section*{Acknowledgments}
KYK and BHL are supported by the Ministry of Education, BK21
Project No. D-0055 and by grant No. R01-1999-00018 from the
Interdisciplinary Research Program of the KOSEF. HSY is supported
by NSC (NSC90-2811-M-002-019). He also acknowledges NCTS as well as
CTP at NTU for partial support.

\newpage


\nc{\np}[3]{Nucl. Phys. {\bf B#1}, #2 (#3)}

\nc{\pl}[3]{Phys. Lett. {\bf B#1}, #2 (#3)}

\nc{\prl}[3]{Phys. Rev. Lett. {\bf #1}, #2 (#3)}

\nc{\prd}[3]{Phys. Rev. {\bf D#1}, #2 (#3)}

\nc{\ap}[3]{Ann. Phys. {\bf #1}, #2 (#3)}

\nc{\prep}[3]{Phys. Rep. {\bf #1}, #2 (#3)}

\nc{\ptp}[3]{Prog. Theor. Phys. {\bf #1}, #2 (#3)}

\nc{\rmp}[3]{Rev. Mod. Phys. {\bf #1}, #2 (#3)}

\nc{\cmp}[3]{Comm. Math. Phys. {\bf #1}, #2 (#3)}

\nc{\mpl}[3]{Mod. Phys. Lett. {\bf #1}, #2 (#3)}

\nc{\cqg}[3]{Class. Quant. Grav. {\bf #1}, #2 (#3)}

\nc{\jhep}[3]{J. High Energy Phys. {\bf #1}, #2 (#3)}

\nc{\hep}[1]{{\tt hep-th/{#1}}}


\end{document}